\documentclass{article}
\usepackage{pdfpages}
\usepackage{float}
\usepackage{bbm}
\usepackage{amssymb}
\usepackage{multicol}
\usepackage{amsmath}
\usepackage{authblk}
\usepackage{algorithm}
\usepackage{amsfonts}
\usepackage{graphicx}
\usepackage{booktabs}
\usepackage{caption}
\usepackage{siunitx}
\usepackage{multirow}
\usepackage{helvet}
\usepackage[T1]{fontenc}
\usepackage[colorlinks=true, allcolors=red]{hyperref}

\DeclareUnicodeCharacter{2212}{-}

\usepackage[a4paper,top=2.5cm,bottom=2.5cm,left=1.3cm,right=1.3cm,marginparwidth=1.75cm]{geometry}

\usepackage[
    backend=biber,
    style=nature,
    natbib=true,
    url=false,
    doi=true
  ]{biblatex}

\AtEveryBibitem{
\clearfield{month}
\clearfield{day}
\clearlist{language}
\clearfield{issn}
\clearlist{publisher}
\clearlist{extra}
\clearfield{note}
\clearname{editor}
}

\title{Learning continually with representational drift}

\author[1]{Suzanne van der Veldt}
\author[2]{Gido M.\ van de Ven}
\author[1]{Sanne Moorman}
\author[1,*]{Guillaume Etter}

\affil[1]{Groningen Institute for Evolutionary Life Sciences, University of Groningen}
\affil[2]{Bernoulli Institute, University of Groningen}
\affil[*]{Correspondence: {g.etter}@rug.nl}
\date{
  \parbox{\linewidth}{\centering%
  \today\endgraf\bigskip}
  }

\begin{document}
\maketitle

\begin{abstract}
Deep artificial neural networks famously struggle to learn from non-stationary streams of data.
Without dedicated mitigation strategies, continual learning is associated with continuous forgetting of previous tasks and a progressive loss of plasticity.
Current approaches to continual learning have either focused on increasing the stability of representations of past tasks, or on promoting plasticity for future learning.
Paradoxically, while animals including humans achieve a desirable stability-plasticity trade-off, the responses of biological neurons to external stimuli that are associated with stable behaviors gradually change over time.
This suggests that, although unstable representations have historically been seen as undesirable in artificial systems, they could be a core property of biological neural networks learning continually.
Here, we examine how linking representational drift to continual learning in biological neural networks could inform artificial systems.
We highlight the existence of representational drift across numerous animal species and brain regions and propose that drift reflects a mixture of homeostatic turnover and learning-related synaptic plasticity.
In particular, we evaluate how plasticity induced by learning new tasks could induce drift in the representation of previous tasks, and how such drift could accumulate across brain regions.
In deep artificial neural networks, we propose that representational drift is only compatible with approaches that do not explicitly prevent parameter changes to mitigate forgetting.
Remarkably, jointly promoting plasticity while mitigating forgetting could in principle induce representational drift in continual learning.
While we argue that drift is a byproduct rather than a solution to incremental learning, its investigation could inform approaches to continual learning in artificial systems. 
\end{abstract}

\begin{multicols}{2}
\raggedcolumns
\section*{Main}
Deep artificial neural networks (ANNs) are currently the major framework in artificial intelligence as they increasingly demonstrate human or above-human performance on numerous tasks \cite{silver_mastering_2016, jumper_highly_2021}.
These models take direct inspiration from biological neural networks (BNNs) by explicitly implementing neurons as units of computation, synapses as learnable parameters, non-linear activation functions, and deep architectures that act as universal function approximators \cite{lecun_deep_2015}.
However, a long-standing limitation of ANNs is their inability to accumulate knowledge continually i.e. to learn from continuous streams of data rather than from large datasets all at once.
In absence of any dedicated regularizing strategy in ANNs, learning new tasks is associated with large changes in representations and rapid forgetting of previous tasks \cite{mccloskey1989catastrophic,ratcliff1990connectionist,french_catastrophic_1999, van_de_ven_three_2022}.
More recently, standard ANNs were also shown to lose plasticity over the course of training, gradually decreasing their ability to learn new information \cite{dohare_loss_2024}.
Currently, approaches to continual learning of ANNs focus on mitigating one limitation at a time, with some combating catastrophic forgetting, and others loss of plasticity.

\begin{figure}[H]
\includegraphics[width=\columnwidth]{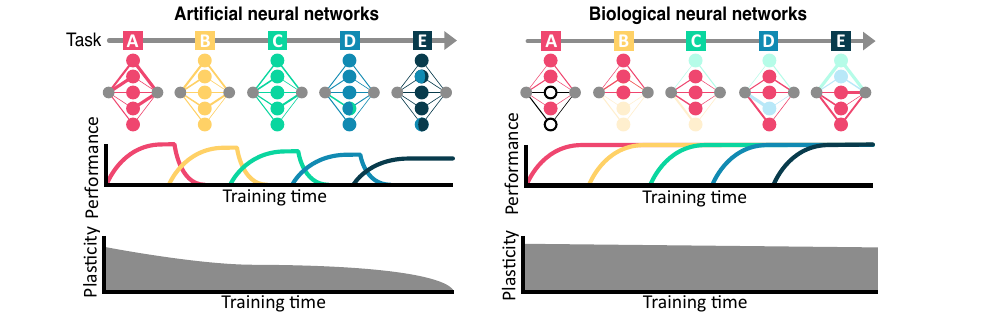}
  \captionsetup{width=\columnwidth}
  \caption{
    \textbf{Linking representational drift and continual learning}.
    Unlike biological systems, standard deep ANNs (left) exhibit catastrophic forgetting and loss of plasticity when learning continually.
    Intriguingly, in BNNs (right), neural representations of previously learned tasks gradually change over time while supporting stable performance.
    Linking such representational drift to learning mechanisms could help define principles of continual learning in BNNs and inform artificial systems.
}
  \label{fig:fig_TLDR}
\end{figure}

Paradoxically, while BNNs are reasonably well suited for learning continually, they are supported by a highly unstable hardware: synapses undergo constant homeostatic turnover \cite{stettler_axons_2006,holtmaat_experience-dependent_2009,trachtenberg_long-term_2002, grutzendler_long-term_2002} and neural representations continuously change in spite of supporting stable behavioral performance \cite{driscoll_representational_2022, rule_causes_2019,micou_representational_2023, ziv_long-term_2013, khatib_active_2023, geva_time_2023,devalle_representational_2025}.
These changes in representations associated with stable performance have been termed `representational drift'.
This intriguing instability in BNNs prompts to evaluate the potential link between representational drift and continual learning in neural networks (Fig.~\ref{fig:fig_TLDR}).

Here, we highlight the pervasiveness of representational drift across species and brain regions.
We examine plausible plasticity mechanisms involved in the gradual drift of representations.
We hypothesize that representational drift reflects plasticity processes involved in maintaining performance under changes induced by homeostatic turnover and plasticity related to learning new tasks.
In ANNs, we evaluate the compatibility of current continual learning approaches with representational drift.
In particular, approaches that do not explicitly prevent changes in neural networks are in principle compatible with drifting representations.
Crucially, we discuss that recent approaches for mitigating loss of plasticity in ANNs implement a form of synaptic turnover, which actively drives changes in the representation of past tasks during the learning of new ones.
We propose that drift itself does not improve continual learning capabilities, but could rather be the byproduct of neural networks that effectively balance stability and plasticity.

\end{multicols}

\begin{multicols}{2}
\raggedcolumns

\section*{Representational drift in biological neural networks}
Historically, neurophysiological studies have characterized neural representations through tuning curves, which describe the selectivity of individual neurons to specific variables, including the sensory \cite{hubel_receptive_1962}, spatial \cite{okeefe_place_1976} or motor \cite{georgopoulos_neuronal_1986} domains.
At shorter timescales, the apparent stability of these representations can fluctuate due to intrinsic mechanisms, such as changes in synaptic release probability~\cite{branco_probability_2009} or transient structural plasticity of dendritic spines and axonal boutons \cite{holtmaat_experience-dependent_2009}.
Recent advances in optical imaging techniques, including calcium imaging of genetically expressed fluorescent sensors \cite{grienberger_imaging_2012, ghosh_miniaturized_2011, stosiek_vivo_2003, cai_shared_2016, aharoni_circuit_2019}, as well as in computational tools to track neuron identities over long periods of time \cite{sheintuch_tracking_2017}, have enabled longitudinal recordings of the same neurons in awake, behaving animals across periods ranging from weeks to several months.
Together, these methods now allow the direct investigation of how neuronal representations evolve over extended timescales under constant stimulus or task conditions.

\subsection*{Evidence for representational drift}
\begin{figure}[H]
\includegraphics[width=\columnwidth]{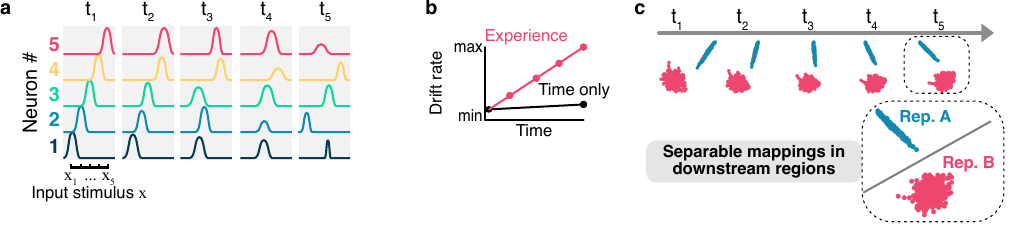}
  \captionsetup{width=\columnwidth}
  \caption{
    \textbf{Representational drift in biological neural networks}.
    \textbf{a}, representational drift refers to gradual, non-mean-reverting changes in how individual neurons encode constant inputs over time, without noticeable changes in behavioral outputs.
    \textbf{b}, representational drift increases with time, but its rate accelerates with experience (successive presentation to the same task).
    \textbf{c}, in spite of representational drift, the structure of high-dimensional representations can remain stable and may remain separable by downstream neurons.
}
  \label{fig:fig_drift}
\end{figure}

Strikingly, recent longitudinal studies report non-rigid representations in biological systems, where the tuning of individual neurons gradually changes in the absence of experimental changes \cite{ahmed_representational_2024, bauer_sensory_2024, deitch_representational_2021, driscoll_dynamic_2017,geva_time_2023, keinath_representation_2022, marks_stimulus-dependent_2021, morales_representational_2025, rokni_motor_2007, roth_representations_2023, rule_stable_2020, schoonover_representational_2021, ziv_long-term_2013}.
This gradual change in how individual neurons encode the same stimulus, context or behavior over time under stable external and behavioral conditions has been termed `representational drift' \cite{rule_causes_2019, driscoll_representational_2022} (Fig.~\ref{fig:fig_drift}\textbf{a}).
Such drift unfolds across a range of timescales, from subtle changes over minutes or hours \cite{aitken_geometry_2022, deitch_representational_2021, khatib_active_2023}, to cumulative transformations across days or weeks \cite{ziv_long-term_2013, keinath_representation_2022, geva_time_2023, khatib_active_2023}.

While time alone can induce representational drift as shown by having animals explore the same environment at distinct time points~\cite{ziv_long-term_2013}, \citet{khatib_active_2023} found that regularly re-exposing animals to the same environment accelerated the rate of drift in the hippocampus (Fig.~\ref{fig:fig_drift}\textbf{b}).
These two experimental conditions have been referred to as `time' versus `experience', and other studies support this phenomenon~\cite{geva_time_2023,rokni_motor_2007, driscoll_dynamic_2017}.
Remarkably, despite drift at the level of individual neurons, the information that can be read out from large populations of neurons often remains comparatively stable \cite{rule_stable_2020, gallego_long-term_2020, xia_stable_2021, keinath_representation_2022} (Fig.~\ref{fig:fig_drift}\textbf{c}).

\end{multicols}
\begin{multicols}{2}
\raggedcolumns

\subsection*{Representational drift is pervasive across biological systems}
The existence of representational drift has previously been questioned based on the observation of stable representations in specific conditions.
For instance, bats navigating fixed flight routes maintain exceptionally stable spatial tuning~\cite{liberti_stable_2022}, and in non-human primates, face-patch neurons can preserve selectivity for months~\cite{mcmahon_face-selective_2014}.
Songbirds provide a particularly illustrative case: Zebra finches repeatedly sing a single, well-learned song with minimal variation, driven by the precisely timed firing of neurons in the premotor nucleus HVC~\cite{burke_neural_2020, aronov_two_2011, hahnloser_ultra-sparse_2002, markowitz_mesoscopic_2015, liberti_unstable_2016, katlowitz_stable_2018}.
Despite this stability, neural representations in HVC can drift over time, especially during undirected `practice' singing or after periods of sleep, when plasticity and learning are most active~\cite{liberti_unstable_2016}.
In contrast, during stereotyped, performance-mode singing, drift is minimal~\cite{katlowitz_stable_2018}.

On the other hand, representational drift has been reported across numerous species, including rodents~\cite{ziv_long-term_2013, keinath_representation_2022, rubin_revealing_2019, kentros_increased_2004, muzzio_attention_2009, kennedy_motivational_2009,pettit_hippocampal_2022, thompson_long-term_1990, elyasaf_novel_2024}, non-human primates  \cite{chestek_single-neuron_2007, rokni_motor_2007}, songbirds~\cite{liberti_unstable_2016, katlowitz_stable_2018}, humans~\cite{roth_representations_2023, rait_hippocampal_2025} and even zebrafish~\cite{jacobson_experience-dependent_2018}, suggesting that this phenomenon is not restricted to mammals but may represent a general property of vertebrate neural networks. 

Altogether, these studies suggest that representational drift is a highly conserved property of BNNs, but it is not a universal rule as representations can sometimes be exceptionally stable.
In the following section, we specifically address inter-regional differences in drift rates.

\subsection*{Circuit-level representational drift}
\begin{figure}[H]
\includegraphics[width=\columnwidth]{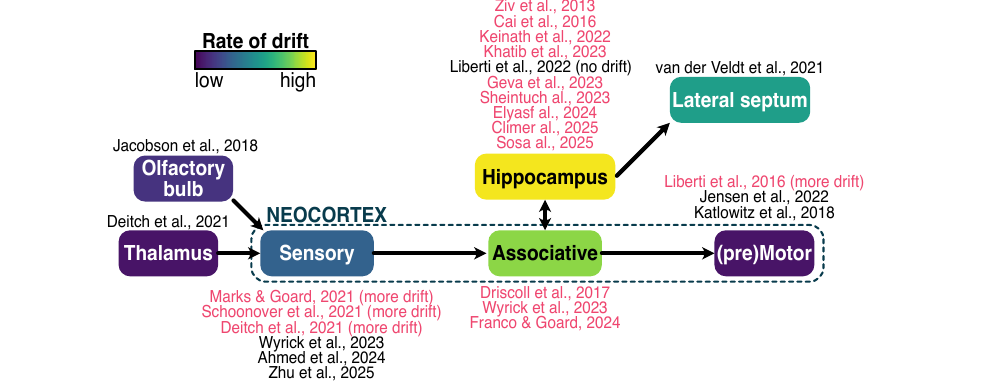}
  \captionsetup{width=\columnwidth}
  \caption{
    \textbf{Representational drift across hierarchical levels.}
    Rate of drift in sensory, hippocampus and associative regions (including the retrosplenial cortex), as well as motor-related (including premotor) regions.
    Dark blue, highly stable representations; yellow, high rate of representational drift.
    References are color-coded with respect of their main finding (red, high drift; black, low/no drift).
    Note that this is a broad summary, as some studies found a mix between representational drift and stability depending on other specific parameters including behavioral tasks.
}
  \label{fig:fig_drift_properties}
\end{figure}

Drift has been described in numerous brain regions, with varying degrees of stability.
First, regions involved in early sensory processing such as the olfactory bulb \cite{jacobson_experience-dependent_2018}, thalamus \cite{deitch_representational_2021}, and low-order sensory cortex \cite{wyrick_differential_2023, zhu_temporal_2025, marks_stimulus-dependent_2021, ahmed_representational_2024, deitch_representational_2021, wyrick_differential_2023, zhu_temporal_2025} displayed slower rates of drift.
Second, regions specialized for rapid learning and episodic encoding, most prominently the hippocampus \cite{keinath_representation_2022, khatib_active_2023, etter_optogenetic_2023, ziv_long-term_2013, rubin_revealing_2019, sheintuch_organization_2023, climer_hippocampal_2025} and cortical associative regions \cite{franco_differential_2024, wyrick_differential_2023, driscoll_dynamic_2017}, show pronounced and continuous drift in their neural representations, with rates that increase with learning and experience \cite{geva_time_2023, khatib_active_2023}.
Subcortical lateral septum neurons downstream of the hippocampus show comparatively more stable coding \cite{van_der_veldt_conjunctive_2021, etter_idiothetic_2024} (Fig.~\ref{fig:fig_drift_properties}).
Finally, regions involved with generating motor outputs often display slower rates of drift \cite{jensen_long-term_2022, katlowitz_stable_2018} (but see \cite{liberti_unstable_2016}).
Similarly, neocortical representations tied to long-term knowledge are relatively stable over comparable timescales \cite{mcmahon_face-selective_2014}.

Together, these observations suggest that drift rates follow an organization that is not strictly hierarchical in nature~\cite{deitch_representational_2021}.
Instead, it is maximal in the hippocampus and minimal in regions involved in either sensory or motor processing (Fig.~\ref{fig:fig_drift_properties}).
It is noteworthy that in the mammalian neocortex and associated regions, neuroanatomical connectivity is rarely strictly feedforward, and higher-order cortical regions can project directly to lower-order sensory or motor regions \cite{harris_hierarchical_2019,felleman_distributed_1991}.
This neuroanatomical feature could explain why drift is still relatively low in motor regions, whereas it would be expected to be maximal in the case of a strictly feed-forward neural network.
Therefore, drift is likely to be additive such that every hierarchical level away from stable sensory or motor regions accumulates drift by adding local plasticity to its more stable, inherited representations.

This `map of drift' also highlights an interesting property of the hippocampus: it is both highly unstable in its representations and neuroanatomically deep.
This is remarkable as the hippocampus has long been known for its role in maintaining recent, but not distant, memories, as supported by early lesion studies \cite{scoville_loss_1957, squire_memory_1992, mcclelland_why_1995, frankland_organization_2005}.
More broadly, this idea is a major tenet of the complementary learning systems theory \cite{mcclelland_why_1995}, the hippocampus supports rapid learning of in-context episodic experiences, while the neocortex slowly extracts general context-invariant representations and latent semantic structure \cite{mcclelland_why_1995, oreilly_complementary_2014, kumaran_what_2016}.

\end{multicols}

\begin{multicols}{2}
\raggedcolumns

\section*{An integrative account of representational drift in neural networks}

Representational drift is a phenomenon that has been described recently and remains under active investigation.
In this section, based on the current literature, we attempt to formalize some key defining characteristics of representational drift, to distinguish it from other related phenomena, and to identify several plausible sources.
One objective hereby is to facilitate establishing links between representational drift in BNNs and continual learning in ANNs.

\subsection*{Distinguishing representational drift from other phenomena}
\begin{figure}[H]
\includegraphics[width=\columnwidth]{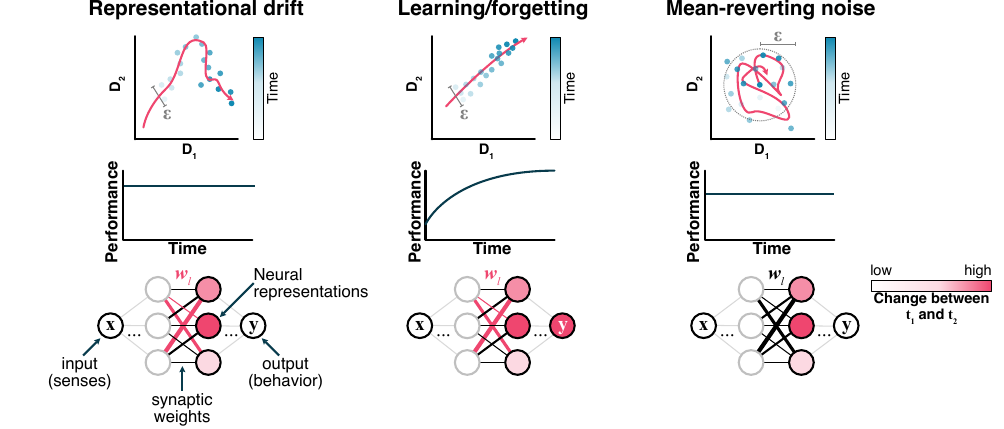}
  \captionsetup{width=\columnwidth}
  \caption{
    \textbf{Representational drift as plasticity without learning.}
    We formalize representational drift as non-mean-reverting changes in neural representations of a task (top, representing a low-dimensional projection of a high-dimensional representation) that are not associated with changes in performance on that task (middle), sensory inputs or motor outputs (bottom).
    This way, representational drift can be contrasted to learning and forgetting that are associated with changes in neural representations, performance, and synaptic weights.
    Finally, representational changes can be induced by mean-reverting processes including changes in excitability, attentional drive, or stochastic synaptic release.
    This noise $\epsilon$ can also be present in the first two scenarios.
}
  \label{fig:fig_drift_definition}
\end{figure}

In the neuroscience literature, representational drift is typically described as changes to neuronal responses in the absence of changes to experimental conditions~\cite{micou_representational_2023}.
We formalize this in a way that is consistent with the literature on continual learning by proposing that representational drift corresponds to changes in neural representations of a task, without changes in performance for that task.
Here, `task' refers to a group of inputs over which a biological or artificial system has to make decisions: for example, a collection of images to classify, an environment comprised of distinct spatial locations, or a set of odors to discriminate from.
Crucially, representational drift should be distinguished from cases where performance changes (improves or deteriorates).
We will refer to this second scenario as `learning or forgetting'.
A core element of discussion that we will develop in the next section is that plasticity can take place in absence of performance changes.
Finally, a key property of representational drift is that it is not mean-reverting i.e. changes are cumulative over time and do not eventually revert to an original state (Fig.~\ref{fig:fig_drift},~\ref{fig:fig_drift_definition}).
Mean-reverting noise can be caused by a variety of factors, including variability in cellular excitability, the stochastic nature of synaptic release, or the internal state of neural network including attentional priors \cite{micou_representational_2023}.
Note that we do not discount mean-reverting noise to occur in conjunction to representational drift.
If such noise is present during learning and plasticity, it could also contribute to cumulative drift over time.
As pointed out by \citet{micou_representational_2023}, one important caveat with this definition is that it assumes ideal conditions where sensory inputs and behavioral outputs remain perfectly identical between two time points.
This is rarely the case in biological systems and might account for some variability in representations.
Thus, experimentally, representational drift is best defined over long periods of time and using as many sampling epochs as possible.

We can briefly formalize drift by considering a simple neural network $f$ with parameters $\theta$ that include synaptic weights $W$.
This neural network processes inputs $x$ (e.g.,~senses in the biological case) and learns a mapping to outputs $y$ (e.g.,~behavior, decisions) by propagating representations $h_l$ at each hierarchical level $l$ using $h_l = h_{l-1}*w_l + \epsilon$, where $w_l$ refers to presynaptic weights and $\epsilon$ to mean-reverting noise.
In this context, representational drift refers specifically to the case where changes in synaptic weights $W$ that lead to distinct task representations $h_{l}$, in absence of changes in inputs $x$ and output $y$ so that $\Delta \mathbf{h}_{l}>0$, with $\Delta \mathbf{h}$ representing the difference in representations at two distinct time points (Fig.~\ref{fig:fig_drift_definition}).

Drift can thus be seen as the result of plasticity mechanisms that lead to changes in neural representation of a task in absence of performance changes for that task.

\end{multicols}

\begin{multicols}{2}
\raggedcolumns

\subsection*{Plausible sources of representational drift}
While drift is most plausibly induced by plasticity, two distinct phenomena may be involved: homeostatic processes, and synaptic plasticity.
Homeostatic processes can either take place at the single neuron level (e.g.,~long-lasting changes in excitability) or at the level of synapses between neurons, and lead to permanent changes in neuronal responses.
In the brain, synapses undergo constant turnover over time in absence of any dedicated learning signal \cite{stettler_axons_2006,holtmaat_experience-dependent_2009,trachtenberg_long-term_2002, grutzendler_long-term_2002}.
This process can be implemented in artificial neural networks by pulling or resetting synaptic weights towards their randomly initialized values~\cite{dohare_loss_2024}, and we will discuss implications for this particular case in a following section.
Crucially, in absence of other plasticity processes, cumulative turnover would lead to forgetting, as learned weights would gradually revert to their initial state.
For this reason, homeostatic synaptic turnover can be seen as a contributor to representational drift, but not the sole driver.

\begin{figure}[H]
\includegraphics[width=\columnwidth]{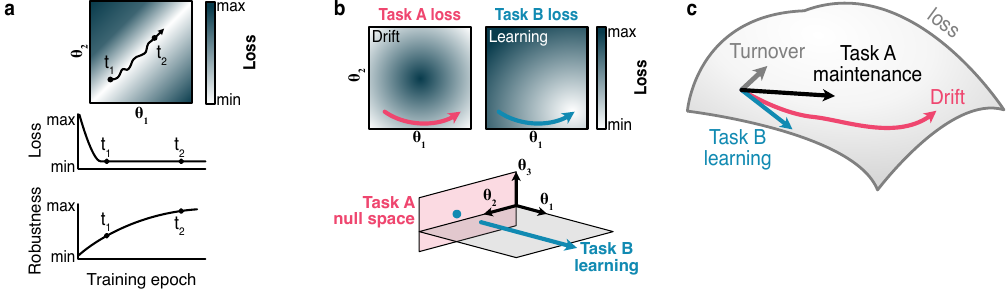}
  \captionsetup{width=\columnwidth}
  \caption{
    \textbf{Plausible sources of representational drift}.
    \textbf{a}, representational drift may reflect plasticity processes associated with learning in a performance plateau.
    This diffusion process may lead neural networks to find solutions that are more robust to weight perturbations \cite{natrajan_stability_2025}.
    \textbf{b}, learning a new task B could induce plasticity and changes in representations of previous task A (in absence of performance changes; top).
    Alternatively, learning task B could take place in the nullspace of task A representations (bottom).
    \textbf{c}, three main forces can drive representational drift.
    Homeostatic processes including synaptic turnover pull representations toward their initial state (gray).
    Synaptic plasticity associated with learning new tasks (blue) could also lead to forgetting of task A.
    Plasticity involved with actively maintaining performance on task A gradually shift representations of task A with while preserving performance (red).
}
  \label{fig:fig_multitask_drift}
\end{figure}

We can contrast these homeostatic processes with synaptic plasticity, which refers to small updates (strengthening or weakening) to connections between neurons.
As proposed in the previous section, changes in representations associated with learning and forgetting -- as defined by performance improvements and deterioration, respectively -- are not compatible with a definition of drift.
On the other hand, if performance remains stable, drift can be seen as an exploration of distinct configurations associated with optimal, plateau performance (or a flat loss landscape in the case of ANNs) \cite{driscoll_representational_2022, micou_representational_2023}.
This diffusion process may have some benefits as neural networks could explore more robust solutions involving sparser representations with larger synaptic weights \cite{natrajan_stability_2025} (Fig.~\ref{fig:fig_multitask_drift}\textbf{a}).

While the vast majority of studies on representational drift focus on a single task, the effects of learning new tasks on previous representations is of high interest for continual learning, as they may induce drift on previously learned representations.
Two cases can be examined: first, learning of new tasks is associated with updates constrained to the nullspace of previous representations (Fig. \ref{fig:fig_multitask_drift}\textbf{b}).
In this case, representations of new tasks are orthogonal to those previously learned, so no maintenance-related plasticity processes are required to prevent forgetting previous tasks.

On the other hand, learning to improve performance on new tasks could also be associated with the gradual forgetting of previous tasks \cite{french_catastrophic_1999}.
In this case, drift could reflect plasticity mechanisms involved with the \textit{active maintenance} of previously learned representations.
Beyond drift, such active maintenance could also take place in parallel to passive mechanisms, including strengthening learned synaptic weights \cite{redondo_making_2011}, which has inspired some approaches to continual learning \cite{zenke_continual_nodate}.
Note that both nullspace learning and protecting important synapses have been leveraged in the machine learning setting and will be discussed in a later section.

To our knowledge, only a few studies directly investigated the effects of learning new tasks on the representational drift.
Of interest is \citet{pashakhanloo_contribution_2025} suggesting that in deep ANNs, learning additional tasks could accelerate the rate of drift of previous task representations.
Surprisingly, \citet{elyasaf_novel_2024} found opposing results in the biological setting: exposing mice to new environments (which could be considered as new tasks) actually slowed down the rate of drift for representations of a previous, reference task.
Further investigation in this direction could shed light on principles used in the brain to accumulate knowledge incrementally.

\end{multicols}

\begin{multicols}{2}
\raggedcolumns

\section*{Linking drift and continual learning in artificial neural networks}

In this section, we discuss how compatible existing approaches for continual learning with ANNs are with representational drift, we hypothesize about a potential role for drift-related processes in alleviating loss of plasticity, and we speculate about how plasticity and stability could be combined in ANNs.

\begin{table*}[b]
    \centering
    \begin{tabular}{lll}
        \toprule
        \multicolumn{2}{c}{\bf Stability-focused~~~~~~~~~~~~~~~} & \bf Plasticity-focused \\
        \midrule
        Task-specific components  & Orthogonal optimization  & Reset weights  \\
        Parameter regularization & Replay / functional regularization   & Inject diversity  \\
        \midrule
        \it Incompatible with drift & \it Allow for drift & \it Potential to induce drift \\
        \bottomrule
    \end{tabular}
    \caption{Overview of approaches for continual learning with ANNs and their relation to representational drift.}
    \label{tab:cl_methods}
\end{table*}

\subsection*{Compatibility of existing stability-focused continual learning approaches with drift}

When an ANN is trained on a sequence of tasks without a dedicated continual learning strategy, learning a new task generally leads to large changes in the network's activations in response to previous tasks. These shifting representations are associated with `catastrophic forgetting' of past tasks~\cite{mccloskey1989catastrophic,ratcliff1990connectionist}. Since in this case, performance is not stable, these changes in activations do not qualify as representational drift. To prevent or alleviate catastrophic forgetting, various continual learning approaches have been proposed (for reviews, see~\cite{hadsell_embracing_2020,delange2021continual,VANDEVEN2025153}). In the following, we discuss to what extent these different approaches exhibit representational drift.

When training on a new task, to promote stability, many continual learning approaches prevent or minimize changes to specific parts of the network.
For example, after finishing training on a new task, methods such as progressive neural networks~\cite{rusu2016progressive} or dynamically expandable networks~\cite{yoon_lifelong_2018}, freeze the part of the network that learned that task, such that it can no longer be updated when learning new tasks.
Another way in which specialized sub-networks have been designed is by gating the network's parameters differently for each task~\cite{masse2018alleviating,serra2018overcoming}.
The use of such task-specific components can guarantee stable behavior on past tasks, but also implies that the representations for previously learned tasks will not change during continued training.

Parameter regularization is another continual learning approach that tries to achieve stability by discouraging large changes to a network. When training on a new task, this approach adds a term to the loss function that penalizes updates of synaptic weights that are associated with solving previous tasks \cite{kirkpatrick_overcoming_2017, zenke_continual_nodate, ritter_online_2018, aljundi_memory_2018, chaudhry_riemannian_2018, liu_rotate_2018, lee_continual_2020, park_continual_2019, benzing_unifying_2022}. Intuitively, it seems that such an explicit penalty on changes to important parameters should actively counter changes to representations of past tasks. \citet{kao2021natural} empirically showed that neural responses to previously learned tasks are indeed rather stable when an ANN continually learns new tasks using parameter regularization.

Maintaining stable performance on past tasks can also be achieved by forcing ANNs to learn new tasks in the null space or orthogonal complement of past tasks~\cite{zeng2019continual,farajtabar2020orthogonal,saha2021gradient}. Such orthogonal optimization can be achieved by restricting parameter updates to directions that do not interfere with the performance on past tasks, for example using gradient projection. As discussed above, in the computational neuroscience literature, it has been hypothesized that continued learning in the null space of past tasks could lead to representational drift for past tasks~\cite{rule_causes_2019} (Fig.~\ref{fig:fig_multitask_drift}\textbf{b}). In ANNs, \citet{anthes2024continual} empirically verified that when using a continual learning method based on orthogonal optimization, neuronal responses to previous tasks can indeed change while performance on those tasks remains stable.

Another continual learning approach that could be compatible with drift is replay~\cite{robins1995catastrophic,shin_continual_2017,van2018generative,rolnick_experience_2019,chaudhry_tiny_2019,buzzega2020dark}.
The idea is that if, due to learning new tasks, in a particular layer the representation of a past task changes, replay allows downstream layers to adjust and to `track' the drift in upstream layers. This way, replay might be able to combine stable performance with drifting representations. Another perspective is that when learning a new task causes interference with previously learned representations, replay allows for correcting them through a form of `active maintenance' (Fig.~\ref{fig:fig_multitask_drift}\textbf{c}). In Appendix~I.4 of~\cite{kao2021natural}, it is shown that with replay, representations of previous tasks can indeed change substantially while maintaining performance on those tasks.
Since functional regularization~\cite{li2017learning,benjamin2019measuring,titsias2020functional,pan2020continual,rudner2022continual} can be interpreted as a form of replay with soft labels~\cite{van2018generative}, we might expect this approach to also allow for representational drift. 
\citet{anthes2024continual} empirically demonstrated that this is indeed the case.

In conclusion, parameter regularization and the use of task-specific components are approaches for continual learning in ANNs that explicitly counteract representational drift. On the other hand, orthogonal optimization, replay and functional regularization are in principle compatible with representational drift (Table \ref{tab:cl_methods}). Importantly however, we argue that drift is not a driver of the performance using these methods: they are \emph{compatible} with drift and \emph{allow for it}, but there is no obvious benefit of drift in terms of stability.

\subsection*{Can drift alleviate loss of plasticity?}

It has recently been pointed out that, in addition to catastrophic forgetting, continual learning in ANNs faces another important issue: loss of plasticity~\cite{dohare_loss_2024}. When an ANN is sequentially trained on many tasks in the standard way, it gradually loses the ability to learn new tasks. One explanation for this is that over the course of continued training, synaptic weights gradually deviate from their initial, maximally plastic state. Given that in computational neuroscience it has been hypothesized that drift is important for enabling the acquisition of new information~\cite{driscoll_representational_2022,rule_causes_2019}, it seems reasonable to ask whether there could be an active role (or `direct benefit') for drift in overcoming loss of plasticity in ANNs.

To counter loss of plasticity, several approaches have been explored, including resetting or pulling weights towards ``randomly initialized values" (e.g., continual backpropagation~\cite{dohare_loss_2024}, weight decay~\cite{kumar2025maintaining}), adding noise or injecting diversity during continued learning (e.g., dropout~\cite{hinton_improving_2012}), or a combination of both (e.g., shrink and perturb~\cite{ash2020warm}).
These approaches have similarities to synaptic turnover, and it seems reasonable to expect that they will induce or enlarge changes in the neuronal responses to past tasks.
Nevertheless, because these plasticity-focused approaches do not prevent forgetting, they cannot be directly related to representational drift as observed in BNNs.

However, we hypothesize that many of these approaches for maintaining plasticity have the \emph{potential} to induce BNN-like representational drift when combined with a suitable stability-focused approach. In particular, we expect the combination of plasticity-based approaches with `active maintenance'-like approaches as replay or functional regularization to be able to lead to both stable performance and continually changing representations. Representational drift might thus be the result of the interplay between plasticity and stability mechanisms.


\section*{Discussion}
Continual learning currently remains one of the most significant challenges for artificial intelligence (AI).
As the demand for AI increases, the energy requirements for repeatedly training large models from scratch pose environmental concerns~\cite{de_vries_growing_2023}, prompting an urgent need for re-usable, continuously trainable AI systems~\cite{verwimp2024continual}.
Insights from biology have informed the design of AI systems capable of continual learning~\cite{kudithipudi_biological_2022,durstewitz2025neuroscience}, but representational drift has largely been overlooked.
From a continual learning perspective, drifting representations are intriguing as they suggest that the stability of some components of neural networks may be relaxed to allow for incremental learning.
Linking representational drift to continual learning in BNNs could be key to further improve the continual learning capabilities of artificial systems.
At the same time, continual learning in ANNs could be a fruitful computational model for representational drift in BNNs~\cite{anthes2024continual}.

One hypothesis we propose here is that representational drift in BNNs may reflect plasticity mechanisms involved in the active maintenance of performance on previously learned tasks.
In particular, a balance between homeostatic synaptic turnover and synaptic plasticity could jointly drive representational drift during task learning at plateau performance.
Several reports support this idea.
Firstly, the rate of homeostatic turnover and synaptic plasticity decrease with age, and should in turn decrease the rate of representational drift, which has been experimentally confirmed by \citet{brown_representational_2025}.
Secondly, experience (successive task presentations) accelerates the rate of drift \cite{khatib_active_2023,geva_time_2023}.
In this case, each exposure to a task could act as a 'training epochs' at plateau performance, and drift could reflect a diffusion process whereby a neural network explores more robust solutions without significantly altering performance \cite{natrajan_stability_2025}.
In BNNs, we do not discount that long-lasting changes may not necessarily involve synaptic plasticity.
As shown by \citet{haimerl_representational_2025}, drift can also be driven by long-term changes in excitability, in absence of synaptic plasticity.

While spatial representations in the bat hippocampus are significantly more stable than in mice \cite{liberti_stable_2022}, it could be speculated that the absence of drift in this case may reflect very low rates of synaptic turnover and plasticity during highly stereotyped behaviors.

We also highlight that drift is minimal (but not absent) in both sensory and motor regions, and maximal in the hippocampus.
Given that in the mammalian brain, both feedforward and feedback propagation of representations are possible \cite{harris_hierarchical_2019,felleman_distributed_1991}, we propose that drift could be accumulated in a hierarchical manner.
As such, drift in the hippocampus would not only be the result of local synaptic processes, but also the accumulation of presynaptic representational drift.
While we highlight that higher drift rates in the hippocampus and lower rates in the neocortex are compatible with the complementary learning system theory \cite{mcclelland_why_1995,mcclelland_considerations_1996}, we also acknowledge that this should be balanced with the observation of intra-regional variability: within the same hierarchical region, some neurons can display more instability that others \cite{sheintuch_organization_2023,franco_differential_2024}.
Several factors could explain this discrepancy.
Most plausibly, projections across brain regions are not homogeneous in the mammalian brain, but depend heavily on the neurochemical identities of neurons \cite{harris_hierarchical_2019}.
Thus, within the same hierarchical level of a BNN, distinct neurons could integrate representations with distinct levels of stability.

Of high interest for continual learning is the effect of learning new tasks sequentially on the drift of past representations.
In one scenario, learning new tasks may take place in the nullspace of previous tasks and does not interfere with representations of known tasks.
On the other hand, learning new tasks could lead to non-orthogonal changes in existing representations and lead to the gradual forgetting of previous tasks.
We posit that in this case, drift might reflect plasticity mechanisms involved in the active maintenance of previous tasks performance.

Experimentally, the effects of continual learning on representational drift remains to be investigated thoroughly.
On the one hand, \citet{pashakhanloo_contribution_2025} shows that in ANNs, drift rates of existing representations accelerate upon the learning of new tasks.
On the other hand, longitudinal recordings of hippocampal neurons suggest that exposure to new tasks (enriched environments) tend to slow down the rate of drift for existing representations \cite{elyasaf_novel_2024}.
More recently, \citet{natrajan_stability_2025} propose that representational drift may be more than `active maintenance' of previous tasks but could reflect the exploration of neural configurations that support the same behavioral outputs but with increased sparsity and robustness to perturbations.

Crucially, key experimental evidence is currently missing.
First, although the effects of time alone versus experience have been explicitly compared \cite{khatib_active_2023}, the impact of continually learning sequential tasks on representational drift remains to be clearly established.
While the influence of exploring novel environments is evaluated in \citet{elyasaf_novel_2024}, tasks that include clear readouts on performance (e.g.,~including a choice) would confirm whether performance remains stable when learning new tasks.
There are only few dedicated investigations of representational drift across neural network hierarchical levels~\cite{deitch_representational_2021}.
While we hypothesize that synaptic turnover (and thus drift) is highest in the hippocampus and supports its function as a short-term buffer, dedicated experiments remain to be carried out to formally define drift rates in pre- and post-synaptic, as well as sensory regions.
While both homeostatic turnover and synaptic plasticity likely contribute to representational drift, the exact balance between these two distinct processes should also be investigated explicitly.
For example, learning can accelerate the rate of turnover in itself~\cite{holtmaat_experience-dependent_2009}, which can be sufficient to alter cellular excitability and induce drift \cite{haimerl_representational_2025}.

Currently, most approaches to continual learning focus on mitigating catastrophic forgetting.
We suggest that not all approaches are compatible with the implementation of a turnover mechanism that could simultaneously promote plasticity in ANNs.
In particular, approaches that actively prevent changes in neural networks (parameter regularization, or freezing task-specific components) seem to directly counteract drift.
On the other hand, we argue that approaches that focus on mitigating loss of plasticity in ANNs~\cite{dohare_loss_2024} have the potential to introduce BNN-like drift in ANNs.
This process has some equivalency to homeostatic turnover, which is coincidentally one of the drivers of representational drift in biological systems.
Synaptic turnover could also introduce additional computational advantages, including sparsifying representations and accelerate training~\cite{malakasis_synaptic_2023}.
Promoting plasticity in ANNs does not mitigate catastrophic forgetting.
However, promoting plasticity while actively preventing forgetting could introduce drift (if the continual learning strategy allows for it).

In conclusion, representational drift may reflect the interplay between processes that promote plasticity and those that actively maintain performance stability.
As such, representational drift provides a unique insight into how a stability-plasticity trade-off is implemented in the brain, and could inform future approaches to continual learning in AI.

\section*{Acknowledgments}
We thank Prof.\ Robbert Havekes and Prof.\ Roelof A. Hut for constructive discussions.

\printbibliography

@article{french_catastrophic_1999,
    title = {Catastrophic forgetting in connectionist networks},
    volume = {3},
    issn = {1364-6613, 1879-307X},
    url = {https://www.cell.com/trends/cognitive-sciences/abstract/S1364-6613(99)01294-2},
    doi = {10.1016/S1364-6613(99)01294-2},
    language = {English},
    number = {4},
    urldate = {2025-05-19},
    journal = {Trends in Cognitive Sciences},
    author = {French, Robert M.},
    month = apr,
    year = {1999},
    pmid = {10322466},
    note = {Publisher: Elsevier},
    pages = {128--135},
}

@article{zenke_continual_nodate,
    title = {Continual {Learning} {Through} {Synaptic} {Intelligence}},
    language = {en},
    author = {Zenke, Friedemann and Poole, Ben and Ganguli, Surya},
}

@article{dohare_loss_2024,
    title = {Loss of plasticity in deep continual learning},
    volume = {632},
    copyright = {2024 The Author(s)},
    issn = {1476-4687},
    url = {https://www.nature.com/articles/s41586-024-07711-7},
    doi = {10.1038/s41586-024-07711-7},
    language = {en},
    number = {8026},
    urldate = {2025-08-01},
    journal = {Nature},
    author = {Dohare, Shibhansh and Hernandez-Garcia, J. Fernando and Lan, Qingfeng and Rahman, Parash and Mahmood, A. Rupam and Sutton, Richard S.},
    month = aug,
    year = {2024},
    note = {Publisher: Nature Publishing Group},
    pages = {768--774},
}

@article{kudithipudi_biological_2022,
    title = {Biological underpinnings for lifelong learning machines},
    volume = {4},
    copyright = {2022 Springer Nature Limited},
    issn = {2522-5839},
    url = {https://www.nature.com/articles/s42256-022-00452-0},
    doi = {10.1038/s42256-022-00452-0},
    language = {en},
    number = {3},
    urldate = {2025-08-18},
    journal = {Nature Machine Intelligence},
    author = {Kudithipudi, Dhireesha and Aguilar-Simon, Mario and Babb, Jonathan and Bazhenov, Maxim and Blackiston, Douglas and Bongard, Josh and Brna, Andrew P. and Chakravarthi Raja, Suraj and Cheney, Nick and Clune, Jeff and Daram, Anurag and Fusi, Stefano and Helfer, Peter and Kay, Leslie and Ketz, Nicholas and Kira, Zsolt and Kolouri, Soheil and Krichmar, Jeffrey L. and Kriegman, Sam and Levin, Michael and Madireddy, Sandeep and Manicka, Santosh and Marjaninejad, Ali and McNaughton, Bruce and Miikkulainen, Risto and Navratilova, Zaneta and Pandit, Tej and Parker, Alice and Pilly, Praveen K. and Risi, Sebastian and Sejnowski, Terrence J. and Soltoggio, Andrea and Soures, Nicholas and Tolias, Andreas S. and Urbina-Meléndez, Darío and Valero-Cuevas, Francisco J. and van de Ven, Gido M. and Vogelstein, Joshua T. and Wang, Felix and Weiss, Ron and Yanguas-Gil, Angel and Zou, Xinyun and Siegelmann, Hava},
    month = mar,
    year = {2022},
    note = {Publisher: Nature Publishing Group},
    pages = {196--210},
}

@article{lecun_deep_2015,
    title = {Deep learning},
    volume = {521},
    copyright = {2015 Springer Nature Limited},
    issn = {1476-4687},
    url = {https://www.nature.com/articles/nature14539},
    doi = {10.1038/nature14539},
    language = {en},
    number = {7553},
    urldate = {2025-05-14},
    journal = {Nature},
    author = {LeCun, Yann and Bengio, Yoshua and Hinton, Geoffrey},
    month = may,
    year = {2015},
    note = {Publisher: Nature Publishing Group},
    pages = {436--444},
}

@misc{hinton_improving_2012,
    title = {Improving neural networks by preventing co-adaptation of feature detectors},
    url = {http://arxiv.org/abs/1207.0580},
    doi = {10.48550/arXiv.1207.0580},
    urldate = {2025-08-19},
    publisher = {arXiv},
    author = {Hinton, Geoffrey E. and Srivastava, Nitish and Krizhevsky, Alex and Sutskever, Ilya and Salakhutdinov, Ruslan R.},
    month = jul,
    year = {2012},
    note = {arXiv:1207.0580 [cs]},
}

@article{kirkpatrick_overcoming_2017,
    title = {Overcoming catastrophic forgetting in neural networks},
    volume = {114},
    issn = {0027-8424, 1091-6490},
    url = {http://arxiv.org/abs/1612.00796},
    doi = {10.1073/pnas.1611835114},
    language = {en},
    number = {13},
    urldate = {2025-08-19},
    journal = {Proceedings of the National Academy of Sciences},
    author = {Kirkpatrick, James and Pascanu, Razvan and Rabinowitz, Neil and Veness, Joel and Desjardins, Guillaume and Rusu, Andrei A. and Milan, Kieran and Quan, John and Ramalho, Tiago and Grabska-Barwinska, Agnieszka and Hassabis, Demis and Clopath, Claudia and Kumaran, Dharshan and Hadsell, Raia},
    month = mar,
    year = {2017},
    note = {arXiv:1612.00796 [cs]},
    pages = {3521--3526},
}

@misc{rolnick_experience_2019,
    title = {Experience {Replay} for {Continual} {Learning}},
    url = {http://arxiv.org/abs/1811.11682},
    doi = {10.48550/arXiv.1811.11682},
    urldate = {2025-08-19},
    publisher = {arXiv},
    author = {Rolnick, David and Ahuja, Arun and Schwarz, Jonathan and Lillicrap, Timothy P. and Wayne, Greg},
    month = nov,
    year = {2019},
    note = {arXiv:1811.11682 [cs]},
}

@article{roth_representations_2023,
    title = {Representations in human primary visual cortex drift over time},
    volume = {14},
    copyright = {2023 This is a U.S. Government work and not under copyright protection in the US; foreign copyright protection may apply},
    issn = {2041-1723},
    url = {https://www.nature.com/articles/s41467-023-40144-w},
    doi = {10.1038/s41467-023-40144-w},
    language = {en},
    number = {1},
    urldate = {2025-08-19},
    journal = {Nature Communications},
    author = {Roth, Zvi N. and Merriam, Elisha P.},
    month = jul,
    year = {2023},
    note = {Publisher: Nature Publishing Group},
    pages = {4422},
}

@article{keinath_representation_2022,
    title = {The representation of context in mouse hippocampus is preserved despite neural drift},
    volume = {13},
    copyright = {2022 The Author(s)},
    issn = {2041-1723},
    url = {https://www.nature.com/articles/s41467-022-30198-7},
    doi = {10.1038/s41467-022-30198-7},
    language = {en},
    number = {1},
    urldate = {2025-08-01},
    journal = {Nature Communications},
    author = {Keinath, Alexandra T. and Mosser, Coralie-Anne and Brandon, Mark P.},
    month = may,
    year = {2022},
    note = {Publisher: Nature Publishing Group},
    pages = {2415},
}

@article{khatib_active_2023,
    title = {Active experience, not time, determines within-day representational drift in dorsal {CA1}},
    volume = {111},
    issn = {0896-6273},
    url = {https://www.cell.com/neuron/abstract/S0896-6273(23)00387-2},
    doi = {10.1016/j.neuron.2023.05.014},
    language = {English},
    number = {15},
    urldate = {2025-08-19},
    journal = {Neuron},
    author = {Khatib, Dorgham and Ratzon, Aviv and Sellevoll, Mariell and Barak, Omri and Morris, Genela and Derdikman, Dori},
    month = aug,
    year = {2023},
    pmid = {37315557},
    note = {Publisher: Elsevier},
    pages = {2348--2356.e4},
}

@article{silver_mastering_2016,
    title = {Mastering the game of {Go} with deep neural networks and tree search},
    volume = {529},
    copyright = {2016 Springer Nature Limited},
    issn = {1476-4687},
    url = {https://www.nature.com/articles/nature16961},
    doi = {10.1038/nature16961},
    language = {en},
    number = {7587},
    urldate = {2025-05-19},
    journal = {Nature},
    author = {Silver, David and Huang, Aja and Maddison, Chris J. and Guez, Arthur and Sifre, Laurent and van den Driessche, George and Schrittwieser, Julian and Antonoglou, Ioannis and Panneershelvam, Veda and Lanctot, Marc and Dieleman, Sander and Grewe, Dominik and Nham, John and Kalchbrenner, Nal and Sutskever, Ilya and Lillicrap, Timothy and Leach, Madeleine and Kavukcuoglu, Koray and Graepel, Thore and Hassabis, Demis},
    month = jan,
    year = {2016},
    note = {Publisher: Nature Publishing Group},
    pages = {484--489},
}

@article{aitken_geometry_2022,
    title = {The geometry of representational drift in natural and artificial neural networks},
    volume = {18},
    issn = {1553-7358},
    url = {https://journals.plos.org/ploscompbiol/article?id=10.1371/journal.pcbi.1010716},
    doi = {10.1371/journal.pcbi.1010716},
    language = {en},
    number = {11},
    urldate = {2025-08-05},
    journal = {PLOS Computational Biology},
    author = {Aitken, Kyle and Garrett, Marina and Olsen, Shawn and Mihalas, Stefan},
    month = nov,
    year = {2022},
    pages = {e1010716},
}

@article{etter_optogenetic_2023,
    title = {Optogenetic frequency scrambling of hippocampal theta oscillations dissociates working memory retrieval from hippocampal spatiotemporal codes},
    volume = {14},
    copyright = {2023 The Author(s)},
    issn = {2041-1723},
    url = {https://www.nature.com/articles/s41467-023-35825-5},
    doi = {10.1038/s41467-023-35825-5},
    language = {en},
    number = {1},
    urldate = {2023-08-07},
    journal = {Nature Communications},
    author = {Etter, Guillaume and van der Veldt, Suzanne and Choi, Jisoo and Williams, Sylvain},
    month = jan,
    year = {2023},
    note = {Number: 1
Publisher: Nature Publishing Group},
    pages = {410},
}

@article{ziv_long-term_2013,
    title = {Long-term dynamics of {CA1} hippocampal place codes},
    volume = {16},
    issn = {1097-6256, 1546-1726},
    url = {http://www.nature.com/articles/nn.3329},
    doi = {10.1038/nn.3329},
    language = {en},
    number = {3},
    urldate = {2019-07-17},
    journal = {Nature Neuroscience},
    author = {Ziv, Yaniv and Burns, Laurie D and Cocker, Eric D and Hamel, Elizabeth O and Ghosh, Kunal K and Kitch, Lacey J and Gamal, Abbas El and Schnitzer, Mark J},
    month = mar,
    year = {2013},
    pages = {264--266},
}

@article{rubin_revealing_2019,
    title = {Revealing neural correlates of behavior without behavioral measurements},
    volume = {10},
    copyright = {2019 The Author(s)},
    issn = {2041-1723},
    url = {https://www.nature.com/articles/s41467-019-12724-2},
    doi = {10.1038/s41467-019-12724-2},
    language = {en},
    number = {1},
    urldate = {2019-12-07},
    journal = {Nature Communications},
    author = {Rubin, Alon and Sheintuch, Liron and Brande-Eilat, Noa and Pinchasof, Or and Rechavi, Yoav and Geva, Nitzan and Ziv, Yaniv},
    month = oct,
    year = {2019},
    pages = {1--14},
}

@article{sheintuch_organization_2023,
    title = {Organization of hippocampal {CA3} into correlated cell assemblies supports a stable spatial code},
    volume = {42},
    issn = {22111247},
    url = {https://linkinghub.elsevier.com/retrieve/pii/S2211124723001304},
    doi = {10.1016/j.celrep.2023.112119},
    language = {en},
    number = {2},
    urldate = {2023-02-21},
    journal = {Cell Reports},
    author = {Sheintuch, Liron and Geva, Nitzan and Deitch, Daniel and Rubin, Alon and Ziv, Yaniv},
    month = feb,
    year = {2023},
    pages = {112119},
}

@article{schoonover_representational_2021,
    title = {Representational drift in primary olfactory cortex},
    volume = {594},
    copyright = {2021 The Author(s), under exclusive licence to Springer Nature Limited},
    issn = {1476-4687},
    url = {https://www.nature.com/articles/s41586-021-03628-7},
    doi = {10.1038/s41586-021-03628-7},
    language = {en},
    number = {7864},
    urldate = {2025-09-18},
    journal = {Nature},
    author = {Schoonover, Carl E. and Ohashi, Sarah N. and Axel, Richard and Fink, Andrew J. P.},
    month = jun,
    year = {2021},
    note = {Publisher: Nature Publishing Group},
    pages = {541--546},
}

@article{marks_stimulus-dependent_2021,
    title = {Stimulus-dependent representational drift in primary visual cortex},
    volume = {12},
    copyright = {2021 The Author(s)},
    issn = {2041-1723},
    url = {https://www.nature.com/articles/s41467-021-25436-3},
    doi = {10.1038/s41467-021-25436-3},
    language = {en},
    number = {1},
    urldate = {2025-09-18},
    journal = {Nature Communications},
    author = {Marks, Tyler D. and Goard, Michael J.},
    month = aug,
    year = {2021},
    note = {Publisher: Nature Publishing Group},
    pages = {5169},
}

@article{xia_stable_2021,
    title = {Stable representation of a naturalistic movie emerges from episodic activity with gain variability},
    volume = {12},
    copyright = {2021 The Author(s)},
    issn = {2041-1723},
    url = {https://www.nature.com/articles/s41467-021-25437-2},
    doi = {10.1038/s41467-021-25437-2},
    language = {en},
    number = {1},
    urldate = {2025-09-18},
    journal = {Nature Communications},
    author = {Xia, Ji and Marks, Tyler D. and Goard, Michael J. and Wessel, Ralf},
    month = aug,
    year = {2021},
    note = {Publisher: Nature Publishing Group},
    pages = {5170},
}

@article{deitch_representational_2021,
    title = {Representational drift in the mouse visual cortex},
    volume = {31},
    issn = {09609822},
    url = {https://linkinghub.elsevier.com/retrieve/pii/S0960982221010526},
    doi = {10.1016/j.cub.2021.07.062},
    language = {en},
    number = {19},
    urldate = {2025-09-18},
    journal = {Current Biology},
    author = {Deitch, Daniel and Rubin, Alon and Ziv, Yaniv},
    month = oct,
    year = {2021},
    pages = {4327--4339.e6},
}

@article{liberti_stable_2022,
    title = {A stable hippocampal code in freely flying bats},
    volume = {604},
    copyright = {2022 The Author(s), under exclusive licence to Springer Nature Limited},
    issn = {1476-4687},
    url = {https://www.nature.com/articles/s41586-022-04560-0},
    doi = {10.1038/s41586-022-04560-0},
    language = {en},
    number = {7904},
    urldate = {2025-09-18},
    journal = {Nature},
    author = {Liberti, William A. and Schmid, Tobias A. and Forli, Angelo and Snyder, Madeleine and Yartsev, Michael M.},
    month = apr,
    year = {2022},
    note = {Publisher: Nature Publishing Group},
    pages = {98--103},
}

@article{jensen_long-term_2022,
    title = {Long-term stability of single neuron activity in the motor system},
    volume = {25},
    copyright = {2022 The Author(s), under exclusive licence to Springer Nature America, Inc.},
    issn = {1546-1726},
    url = {https://www.nature.com/articles/s41593-022-01194-3},
    doi = {10.1038/s41593-022-01194-3},
    language = {en},
    number = {12},
    urldate = {2025-09-18},
    journal = {Nature Neuroscience},
    author = {Jensen, Kristopher T. and Kadmon Harpaz, Naama and Dhawale, Ashesh K. and Wolff, Steffen B. E. and Ölveczky, Bence P.},
    month = dec,
    year = {2022},
    note = {Publisher: Nature Publishing Group},
    pages = {1664--1674},
}

@article{pettit_hippocampal_2022,
    title = {Hippocampal place codes are gated by behavioral engagement},
    volume = {25},
    copyright = {2022 The Author(s)},
    issn = {1546-1726},
    url = {https://www.nature.com/articles/s41593-022-01050-4},
    doi = {10.1038/s41593-022-01050-4},
    language = {en},
    number = {5},
    urldate = {2025-09-18},
    journal = {Nature Neuroscience},
    author = {Pettit, Noah L. and Yuan, Xintong C. and Harvey, Christopher D.},
    month = may,
    year = {2022},
    note = {Publisher: Nature Publishing Group},
    pages = {561--566},
}

@article{kentros_increased_2004,
    title = {Increased {Attention} to {Spatial} {Context} {Increases} {Both} {Place} {Field} {Stability} and {Spatial} {Memory}},
    volume = {42},
    issn = {0896-6273},
    url = {https://www.sciencedirect.com/science/article/pii/S0896627304001928},
    doi = {10.1016/S0896-6273(04)00192-8},
    number = {2},
    urldate = {2025-09-18},
    journal = {Neuron},
    author = {Kentros, Clifford G and Agnihotri, Naveen T and Streater, Samantha and Hawkins, Robert D and Kandel, Eric R},
    month = apr,
    year = {2004},
    pages = {283--295},
}

@article{kennedy_motivational_2009,
    title = {Motivational states activate distinct hippocampal representations to guide goal-directed behaviors},
    volume = {106},
    url = {https://doi.org/10.1073/pnas.0903259106},
    doi = {10.1073/pnas.0903259106},
    number = {26},
    urldate = {2025-09-18},
    journal = {Proceedings of the National Academy of Sciences},
    author = {Kennedy, Pamela J. and Shapiro, Matthew L.},
    month = jun,
    year = {2009},
    note = {Publisher: Proceedings of the National Academy of Sciences},
    pages = {10805--10810},
}

@article{micou_representational_2023,
    title = {Representational drift as a window into neural and behavioural plasticity},
    volume = {81},
    issn = {0959-4388},
    url = {https://www.sciencedirect.com/science/article/pii/S0959438823000715},
    doi = {10.1016/j.conb.2023.102746},
    urldate = {2025-09-18},
    journal = {Current Opinion in Neurobiology},
    author = {Micou, Charles and O'Leary, Timothy},
    month = aug,
    year = {2023},
    pages = {102746},
}

@article{liberti_unstable_2016,
    title = {Unstable neurons underlie a stable learned behavior},
    volume = {19},
    issn = {1546-1726},
    doi = {10.1038/nn.4405},
    language = {eng},
    number = {12},
    journal = {Nature Neuroscience},
    author = {Liberti, William A. and Markowitz, Jeffrey E. and Perkins, L. Nathan and Liberti, Derek C. and Leman, Daniel P. and Guitchounts, Grigori and Velho, Tarciso and Kotton, Darrell N. and Lois, Carlos and Gardner, Timothy J.},
    month = dec,
    year = {2016},
    pmid = {27723744},
    pmcid = {PMC5127780},
    pages = {1665--1671},
}

@article{rule_stable_2020,
    title = {Stable task information from an unstable neural population},
    volume = {9},
    issn = {2050-084X},
    doi = {10.7554/eLife.51121},
    language = {eng},
    journal = {eLife},
    author = {Rule, Michael E. and Loback, Adrianna R. and Raman, Dhruva V. and Driscoll, Laura N. and Harvey, Christopher D. and O'Leary, Timothy},
    month = jul,
    year = {2020},
    pmid = {32660692},
    pmcid = {PMC7392606},
    pages = {e51121},
}

@article{ahmed_representational_2024,
    title = {Representational drift in barrel cortex is receptive field dependent},
    volume = {34},
    issn = {0960-9822},
    url = {https://www.cell.com/current-biology/abstract/S0960-9822(24)01375-7},
    doi = {10.1016/j.cub.2024.10.021},
    language = {English},
    number = {24},
    urldate = {2025-09-30},
    journal = {Current Biology},
    author = {Ahmed, Alisha and Voelcker, Bettina and Peron, Simon},
    month = dec,
    year = {2024},
    pmid = {39541977},
    note = {Publisher: Elsevier},
    pages = {5623--5634.e4},
}

@article{holtmaat_experience-dependent_2009,
    title = {Experience-dependent structural synaptic plasticity in the mammalian brain},
    volume = {10},
    copyright = {2009 Springer Nature Limited},
    issn = {1471-0048},
    url = {https://www.nature.com/articles/nrn2699},
    doi = {10.1038/nrn2699},
    language = {en},
    number = {9},
    urldate = {2025-09-30},
    journal = {Nature Reviews Neuroscience},
    author = {Holtmaat, Anthony and Svoboda, Karel},
    month = sep,
    year = {2009},
    note = {Publisher: Nature Publishing Group},
    pages = {647--658},
}

@article{van_der_veldt_conjunctive_2021,
    title = {Conjunctive spatial and self-motion codes are topographically organized in the {GABAergic} cells of the lateral septum},
    volume = {19},
    issn = {1545-7885},
    url = {https://dx.plos.org/10.1371/journal.pbio.3001383},
    doi = {10.1371/journal.pbio.3001383},
    language = {en},
    number = {8},
    urldate = {2023-05-03},
    journal = {PLOS Biology},
    author = {van der Veldt, Suzanne and Etter, Guillaume and Mosser, Coralie-Anne and Manseau, Frédéric and Williams, Sylvain},
    month = aug,
    year = {2021},
    pages = {e3001383},
}

@article{driscoll_dynamic_2017,
    title = {Dynamic {Reorganization} of {Neuronal} {Activity} {Patterns} in {Parietal} {Cortex}},
    volume = {170},
    issn = {1097-4172},
    doi = {10.1016/j.cell.2017.07.021},
    language = {eng},
    number = {5},
    journal = {Cell},
    author = {Driscoll, Laura N. and Pettit, Noah L. and Minderer, Matthias and Chettih, Selmaan N. and Harvey, Christopher D.},
    month = aug,
    year = {2017},
    pmid = {28823559},
    pmcid = {PMC5718200},
    pages = {986--999.e16},
}

@article{bauer_sensory_2024,
    title = {Sensory experience steers representational drift in mouse visual cortex},
    volume = {15},
    issn = {2041-1723},
    url = {https://pmc.ncbi.nlm.nih.gov/articles/PMC11499870/},
    doi = {10.1038/s41467-024-53326-x},
    urldate = {2025-09-25},
    journal = {Nature Communications},
    author = {Bauer, Joel and Lewin, Uwe and Herbert, Elizabeth and Gjorgjieva, Julijana and Schoonover, Carl E. and Fink, Andrew J. P. and Rose, Tobias and Bonhoeffer, Tobias and Hübener, Mark},
    month = oct,
    year = {2024},
    pmid = {39443498},
    pmcid = {PMC11499870},
    pages = {9153},
}

@article{trachtenberg_long-term_2002,
    title = {Long-term in vivo imaging of experience-dependent synaptic plasticity in adult cortex},
    volume = {420},
    issn = {0028-0836},
    doi = {10.1038/nature01273},
    language = {eng},
    number = {6917},
    journal = {Nature},
    author = {Trachtenberg, Joshua T. and Chen, Brian E. and Knott, Graham W. and Feng, Guoping and Sanes, Joshua R. and Welker, Egbert and Svoboda, Karel},
    month = dec,
    year = {2002},
    pmid = {12490942},
    pages = {788--794},
}

@article{grutzendler_long-term_2002,
    title = {Long-term dendritic spine stability in the adult cortex},
    volume = {420},
    copyright = {2003 Macmillan Magazines Ltd.},
    issn = {1476-4687},
    url = {https://www.nature.com/articles/nature01276},
    doi = {10.1038/nature01276},
    language = {en},
    number = {6917},
    urldate = {2025-09-30},
    journal = {Nature},
    author = {Grutzendler, Jaime and Kasthuri, Narayanan and Gan, Wen-Biao},
    month = dec,
    year = {2002},
    note = {Publisher: Nature Publishing Group},
    pages = {812--816},
}

@article{mcclelland_considerations_1996,
    title = {Considerations arising from a complementary learning systems perspective on hippocampus and neocortex},
    volume = {6},
    issn = {1050-9631},
    doi = {10.1002/(SICI)1098-1063(1996)6:6<654::AID-HIPO8>3.0.CO;2-G},
    language = {eng},
    number = {6},
    journal = {Hippocampus},
    author = {McClelland, J. L. and Goddard, N. H.},
    year = {1996},
    pmid = {9034852},
    pages = {654--665},
}

@article{kumaran_what_2016,
    title = {What {Learning} {Systems} do {Intelligent} {Agents} {Need}? {Complementary} {Learning} {Systems} {Theory} {Updated}},
    volume = {20},
    issn = {1879-307X},
    shorttitle = {What {Learning} {Systems} do {Intelligent} {Agents} {Need}?},
    doi = {10.1016/j.tics.2016.05.004},
    language = {eng},
    number = {7},
    journal = {Trends in Cognitive Sciences},
    author = {Kumaran, Dharshan and Hassabis, Demis and McClelland, James L.},
    month = jul,
    year = {2016},
    pmid = {27315762},
    pages = {512--534},
}

@article{jacobson_experience-dependent_2018,
    title = {Experience-{Dependent} {Plasticity} of {Odor} {Representations} in the {Telencephalon} of {Zebrafish}},
    volume = {28},
    issn = {1879-0445},
    doi = {10.1016/j.cub.2017.11.007},
    language = {eng},
    number = {1},
    journal = {Current biology: CB},
    author = {Jacobson, Gilad A. and Rupprecht, Peter and Friedrich, Rainer W.},
    month = jan,
    year = {2018},
    pmid = {29249662},
    pages = {1--14.e3},
}

@article{rokni_motor_2007,
    title = {Motor learning with unstable neural representations},
    volume = {54},
    issn = {0896-6273},
    doi = {10.1016/j.neuron.2007.04.030},
    language = {eng},
    number = {4},
    journal = {Neuron},
    author = {Rokni, Uri and Richardson, Andrew G. and Bizzi, Emilio and Seung, H. Sebastian},
    month = may,
    year = {2007},
    pmid = {17521576},
    pages = {653--666},
}

@article{elyasaf_novel_2024,
    title = {Novel off-context experience constrains hippocampal representational drift},
    volume = {34},
    issn = {1879-0445},
    doi = {10.1016/j.cub.2024.10.027},
    language = {eng},
    number = {24},
    journal = {Current biology: CB},
    author = {Elyasaf, Gal and Rubin, Alon and Ziv, Yaniv},
    month = dec,
    year = {2024},
    pmid = {39566502},
    pages = {5769--5773.e3},
}

@article{geva_time_2023,
    title = {Time and experience differentially affect distinct aspects of hippocampal representational drift},
    volume = {111},
    issn = {0896-6273},
    url = {https://www.cell.com/neuron/abstract/S0896-6273(23)00378-1},
    doi = {10.1016/j.neuron.2023.05.005},
    language = {English},
    number = {15},
    urldate = {2025-10-06},
    journal = {Neuron},
    author = {Geva, Nitzan and Deitch, Daniel and Rubin, Alon and Ziv, Yaniv},
    month = aug,
    year = {2023},
    pmid = {37315556},
    note = {Publisher: Elsevier},
    pages = {2357--2366.e5},
}

@article{driscoll_representational_2022,
    title = {Representational drift: {Emerging} theories for continual learning and experimental future directions},
    volume = {76},
    issn = {0959-4388},
    shorttitle = {Representational drift},
    url = {https://www.sciencedirect.com/science/article/pii/S0959438822001039},
    doi = {10.1016/j.conb.2022.102609},
    urldate = {2025-10-06},
    journal = {Current Opinion in Neurobiology},
    author = {Driscoll, Laura N. and Duncker, Lea and Harvey, Christopher D.},
    month = oct,
    year = {2022},
    pages = {102609},
}

@article{rule_causes_2019,
    series = {Computational {Neuroscience}},
    title = {Causes and consequences of representational drift},
    volume = {58},
    issn = {0959-4388},
    url = {https://www.sciencedirect.com/science/article/pii/S0959438819300303},
    doi = {10.1016/j.conb.2019.08.005},
    urldate = {2025-08-05},
    journal = {Current Opinion in Neurobiology},
    author = {Rule, Michael E and O’Leary, Timothy and Harvey, Christopher D},
    month = oct,
    year = {2019},
    pages = {141--147},
}

@article{hubel_receptive_1962,
    title = {Receptive fields, binocular interaction and functional architecture in the cat's visual cortex},
    volume = {160},
    issn = {0022-3751},
    url = {https://pmc.ncbi.nlm.nih.gov/articles/PMC1359523/},
    doi = {10.1113/jphysiol.1962.sp006837},
    number = {1},
    urldate = {2025-10-08},
    journal = {The Journal of Physiology},
    author = {Hubel, D. H. and Wiesel, T. N.},
    month = jan,
    year = {1962},
    pmid = {14449617},
    pmcid = {PMC1359523},
    pages = {106--154.2},
}

@article{okeefe_place_1976,
    title = {Place units in the hippocampus of the freely moving rat},
    volume = {51},
    issn = {10902430},
    doi = {10.1016/0014-4886(76)90055-8},
    number = {1},
    journal = {Experimental Neurology},
    author = {O'Keefe, John},
    year = {1976},
    pmid = {1261644},
    note = {ISBN: 0014-4886},
    pages = {78--109},
}

@article{georgopoulos_neuronal_1986,
    title = {Neuronal {Population} {Coding} of {Movement} {Direction}},
    volume = {233},
    url = {https://www.science.org/doi/10.1126/science.3749885},
    doi = {10.1126/science.3749885},
    number = {4771},
    urldate = {2025-10-08},
    journal = {Science},
    author = {Georgopoulos, Apostolos P. and Schwartz, Andrew B. and Kettner, Ronald E.},
    month = sep,
    year = {1986},
    note = {Publisher: American Association for the Advancement of Science},
    pages = {1416--1419},
}

@article{grienberger_imaging_2012,
    title = {Imaging {Calcium} in {Neurons}},
    volume = {73},
    issn = {08966273},
    url = {http://dx.doi.org/10.1016/j.neuron.2012.02.011},
    doi = {10.1016/j.neuron.2012.02.011},
    number = {5},
    journal = {Neuron},
    author = {Grienberger, Christine and Konnerth, Arthur},
    year = {2012},
    pmid = {22405199},
    note = {ISBN: 1097-4199 (Electronic)\${\textbackslash}backslash\$r0896-6273 (Linking)
Publisher: Elsevier Inc.},
    pages = {862--885},
}

@article{ghosh_miniaturized_2011,
    title = {Miniaturized integration of a fluorescence microscope},
    volume = {8},
    issn = {1548-7105},
    doi = {10.1038/nmeth.1694},
    language = {eng},
    number = {10},
    journal = {Nature Methods},
    author = {Ghosh, Kunal K. and Burns, Laurie D. and Cocker, Eric D. and Nimmerjahn, Axel and Ziv, Yaniv and Gamal, Abbas El and Schnitzer, Mark J.},
    month = sep,
    year = {2011},
    pmid = {21909102},
    pmcid = {PMC3810311},
    pages = {871--878},
}

@article{stosiek_vivo_2003,
    title = {In vivo two-photon calcium imaging of neuronal networks},
    volume = {100},
    issn = {0027-8424},
    url = {https://pmc.ncbi.nlm.nih.gov/articles/PMC165873/},
    doi = {10.1073/pnas.1232232100},
    number = {12},
    urldate = {2025-10-08},
    journal = {Proceedings of the National Academy of Sciences of the United States of America},
    author = {Stosiek, Christoph and Garaschuk, Olga and Holthoff, Knut and Konnerth, Arthur},
    month = jun,
    year = {2003},
    pmid = {12777621},
    pmcid = {PMC165873},
    pages = {7319--7324},
}

@article{katlowitz_stable_2018,
    title = {Stable {Sequential} {Activity} {Underlying} the {Maintenance} of a {Precisely} {Executed} {Skilled} {Behavior}},
    volume = {98},
    issn = {1097-4199},
    doi = {10.1016/j.neuron.2018.05.017},
    language = {eng},
    number = {6},
    journal = {Neuron},
    author = {Katlowitz, Kalman A. and Picardo, Michel A. and Long, Michael A.},
    month = jun,
    year = {2018},
    pmid = {29861283},
    pmcid = {PMC6094941},
    pages = {1133--1140.e3},
}

@article{mcmahon_face-selective_2014,
    title = {Face-selective neurons maintain consistent visual responses across months},
    volume = {111},
    url = {https://www.pnas.org/doi/full/10.1073/pnas.1318331111},
    doi = {10.1073/pnas.1318331111},
    number = {22},
    urldate = {2025-10-08},
    journal = {Proceedings of the National Academy of Sciences},
    author = {McMahon, David B. T. and Jones, Adam P. and Bondar, Igor V. and Leopold, David A.},
    month = jun,
    year = {2014},
    note = {Publisher: Proceedings of the National Academy of Sciences},
    pages = {8251--8256},
}

@article{morales_representational_2025,
    title = {Representational drift and learning-induced stabilization in the piriform cortex},
    volume = {122},
    url = {https://www.pnas.org/doi/10.1073/pnas.2501811122},
    doi = {10.1073/pnas.2501811122},
    number = {29},
    urldate = {2025-09-25},
    journal = {Proceedings of the National Academy of Sciences},
    author = {Morales, Guillermo B. and Muñoz, Miguel A. and Tu, Yuhai},
    month = jul,
    year = {2025},
    note = {Publisher: Proceedings of the National Academy of Sciences},
    pages = {e2501811122},
}

@article{gallego_long-term_2020,
    title = {Long-term stability of cortical population dynamics underlying consistent behavior},
    volume = {23},
    copyright = {2020 The Author(s), under exclusive licence to Springer Nature America, Inc.},
    issn = {1546-1726},
    url = {https://www.nature.com/articles/s41593-019-0555-4},
    doi = {10.1038/s41593-019-0555-4},
    language = {en},
    number = {2},
    urldate = {2025-10-08},
    journal = {Nature Neuroscience},
    author = {Gallego, Juan A. and Perich, Matthew G. and Chowdhury, Raeed H. and Solla, Sara A. and Miller, Lee E.},
    month = feb,
    year = {2020},
    note = {Publisher: Nature Publishing Group},
    pages = {260--270},
}

@article{climer_hippocampal_2025,
    title = {Hippocampal representations drift in stable multisensory environments},
    copyright = {2025 The Author(s), under exclusive licence to Springer Nature Limited},
    issn = {1476-4687},
    url = {https://www.nature.com/articles/s41586-025-09245-y},
    doi = {10.1038/s41586-025-09245-y},
    language = {en},
    urldate = {2025-08-01},
    journal = {Nature},
    author = {Climer, Jason R. and Davoudi, Heydar and Oh, Jun Young and Dombeck, Daniel A.},
    month = jul,
    year = {2025},
    note = {Publisher: Nature Publishing Group},
    pages = {1--9},
}

@article{brown_representational_2025,
    title = {Representational drift gates critical-period plasticity in mouse visual cortex},
    volume = {35},
    issn = {1879-0445},
    doi = {10.1016/j.cub.2025.07.026},
    language = {eng},
    number = {17},
    journal = {Current biology: CB},
    author = {Brown, Thomas C. and McGee, Aaron W.},
    month = sep,
    year = {2025},
    pmid = {40769153},
    pmcid = {PMC12419199},
    pages = {4251--4258.e3},
}

@article{etter_idiothetic_2024,
    title = {Idiothetic representations are modulated by availability of sensory inputs and task demands in the hippocampal-septal circuit},
    volume = {43},
    issn = {2211-1247},
    url = {https://www.cell.com/cell-reports/abstract/S2211-1247(24)01331-7},
    doi = {10.1016/j.celrep.2024.114980},
    language = {English},
    number = {11},
    urldate = {2025-08-12},
    journal = {Cell Reports},
    author = {Etter, Guillaume and Veldt, Suzanne van der and Mosser, Coralie-Anne and Hasselmo, Michael E. and Williams, Sylvain},
    month = nov,
    year = {2024},
    pmid = {39535920},
    note = {Publisher: Elsevier},
}

@article{chestek_single-neuron_2007,
    title = {Single-{Neuron} {Stability} during {Repeated} {Reaching} in {Macaque} {Premotor} {Cortex}},
    volume = {27},
    issn = {0270-6474},
    url = {https://pmc.ncbi.nlm.nih.gov/articles/PMC6672821/},
    doi = {10.1523/JNEUROSCI.0959-07.2007},
    number = {40},
    urldate = {2025-10-22},
    journal = {The Journal of Neuroscience},
    author = {Chestek, Cynthia A. and Batista, Aaron P. and Santhanam, Gopal and Yu, Byron M. and Afshar, Afsheen and Cunningham, John P. and Gilja, Vikash and Ryu, Stephen I. and Churchland, Mark M. and Shenoy, Krishna V.},
    month = oct,
    year = {2007},
    pmid = {17913908},
    pmcid = {PMC6672821},
    pages = {10742--10750},
}

@article{rait_hippocampal_2025,
    title = {Hippocampal drift rate reflects the temporal organization of memories},
    copyright = {Copyright © 2025 the authors. SfN exclusive license.},
    issn = {0270-6474, 1529-2401},
    url = {https://www.jneurosci.org/content/early/2025/10/03/JNEUROSCI.0909-25.2025},
    doi = {10.1523/JNEUROSCI.0909-25.2025},
    language = {en},
    urldate = {2025-10-22},
    journal = {Journal of Neuroscience},
    author = {Rait, Lindsay I. and Wanjia, Guo and Ye, Zhifang and DuBrow, Sarah and Kuhl, Brice A.},
    month = oct,
    year = {2025},
    pmid = {41057260},
    note = {Publisher: Society for Neuroscience
Section: Research Articles},
}

@article{muzzio_attention_2009,
    title = {Attention {Enhances} the {Retrieval} and {Stability} of {Visuospatial} and {Olfactory} {Representations} in the {Dorsal} {Hippocampus}},
    volume = {7},
    issn = {1545-7885},
    url = {https://journals.plos.org/plosbiology/article?id=10.1371/journal.pbio.1000140},
    doi = {10.1371/journal.pbio.1000140},
    language = {en},
    number = {6},
    urldate = {2025-10-22},
    journal = {PLOS Biology},
    author = {Muzzio, Isabel A. and Levita, Liat and Kulkarni, Jayant and Monaco, Joseph and Kentros, Clifford and Stead, Matthew and Abbott, Larry F. and Kandel, Eric R.},
    month = jun,
    year = {2009},
    note = {Publisher: Public Library of Science},
    pages = {e1000140},
}

@article{branco_probability_2009,
    title = {The probability of neurotransmitter release: variability and feedback control at single synapses},
    volume = {10},
    copyright = {2009 Springer Nature Limited},
    issn = {1471-0048},
    shorttitle = {The probability of neurotransmitter release},
    url = {https://www.nature.com/articles/nrn2634},
    doi = {10.1038/nrn2634},
    language = {en},
    number = {5},
    urldate = {2025-10-22},
    journal = {Nature Reviews Neuroscience},
    author = {Branco, Tiago and Staras, Kevin},
    month = may,
    year = {2009},
    note = {Publisher: Nature Publishing Group},
    pages = {373--383},
}

@article{cai_shared_2016,
    title = {A shared neural ensemble links distinct contextual memories encoded close in time},
    volume = {534},
    issn = {0028-0836, 1476-4687},
    url = {http://www.nature.com/articles/nature17955},
    doi = {10.1038/nature17955},
    language = {en},
    number = {7605},
    urldate = {2020-03-24},
    journal = {Nature},
    author = {Cai, Denise J. and Aharoni, Daniel and Shuman, Tristan and Shobe, Justin and Biane, Jeremy and Song, Weilin and Wei, Brandon and Veshkini, Michael and La-Vu, Mimi and Lou, Jerry and Flores, Sergio E. and Kim, Isaac and Sano, Yoshitake and Zhou, Miou and Baumgaertel, Karsten and Lavi, Ayal and Kamata, Masakazu and Tuszynski, Mark and Mayford, Mark and Golshani, Peyman and Silva, Alcino J.},
    month = jun,
    year = {2016},
    pages = {115--118},
}

@article{aharoni_circuit_2019,
    title = {Circuit {Investigations} {With} {Open}-{Source} {Miniaturized} {Microscopes}: {Past}, {Present} and {Future}},
    volume = {13},
    issn = {1662-5102},
    shorttitle = {Circuit {Investigations} {With} {Open}-{Source} {Miniaturized} {Microscopes}},
    doi = {10.3389/fncel.2019.00141},
    language = {eng},
    journal = {Frontiers in Cellular Neuroscience},
    author = {Aharoni, Daniel and Hoogland, Tycho M.},
    year = {2019},
    pmid = {31024265},
    pmcid = {PMC6461004},
    pages = {141},
}

@article{thompson_long-term_1990,
    title = {Long-term stability of the place-field activity of single units recorded from the dorsal hippocampus of freely behaving rats},
    volume = {509},
    issn = {0006-8993},
    doi = {10.1016/0006-8993(90)90555-p},
    language = {eng},
    number = {2},
    journal = {Brain Research},
    author = {Thompson, L. T. and Best, P. J.},
    month = feb,
    year = {1990},
    pmid = {2322825},
    pages = {299--308},
}

@article{sheintuch_tracking_2017,
    title = {Tracking the {Same} {Neurons} across {Multiple} {Days} in {Ca2}+{Imaging} {Data}},
    volume = {21},
    issn = {22111247},
    url = {https://doi.org/10.1016/j.celrep.2017.10.013},
    doi = {10.1016/j.celrep.2017.10.013},
    number = {4},
    journal = {Cell Reports},
    author = {Sheintuch, Liron and Rubin, Alon and Brande-Eilat, Noa and Geva, Nitzan and Sadeh, Noa and Pinchasof, Or and Ziv, Yaniv},
    year = {2017},
    pmid = {29069591},
    note = {Publisher: ElsevierCompany.},
    pages = {1102--1115},
}

@misc{chaudhry_tiny_2019,
    title = {On {Tiny} {Episodic} {Memories} in {Continual} {Learning}},
    url = {http://arxiv.org/abs/1902.10486},
    doi = {10.48550/arXiv.1902.10486},
    urldate = {2025-10-21},
    publisher = {arXiv},
    author = {Chaudhry, Arslan and Rohrbach, Marcus and Elhoseiny, Mohamed and Ajanthan, Thalaiyasingam and Dokania, Puneet K. and Torr, Philip H. S. and Ranzato, Marc'Aurelio},
    month = jun,
    year = {2019},
    note = {arXiv:1902.10486 [cs]},
}

@inproceedings{shin_continual_2017,
    title = {Continual {Learning} with {Deep} {Generative} {Replay}},
    volume = {30},
    url = {https://proceedings.neurips.cc/paper/2017/hash/0efbe98067c6c73dba1250d2beaa81f9-Abstract.html},
    urldate = {2025-10-21},
    booktitle = {Advances in {Neural} {Information} {Processing} {Systems}},
    publisher = {Curran Associates, Inc.},
    author = {Shin, Hanul and Lee, Jung Kwon and Kim, Jaehong and Kim, Jiwon},
    doi = {https://doi.org/10.48550/arXiv.1705.08690},
    year = {2017},
}

@inproceedings{ritter_online_2018,
    title = {Online {Structured} {Laplace} {Approximations} for {Overcoming} {Catastrophic} {Forgetting}},
    volume = {31},
    url = {https://proceedings.neurips.cc/paper_files/paper/2018/hash/f31b20466ae89669f9741e047487eb37-Abstract.html},
    urldate = {2025-10-21},
    booktitle = {Advances in {Neural} {Information} {Processing} {Systems}},
    publisher = {Curran Associates, Inc.},
    author = {Ritter, Hippolyt and Botev, Aleksandar and Barber, David},
    doi={https://doi.org/10.48550/arXiv.1805.07810},
    year = {2018},
}

@inproceedings{aljundi_memory_2018,
    title = {Memory {Aware} {Synapses}: {Learning} what (not) to forget},
    shorttitle = {Memory {Aware} {Synapses}},
    url = {https://openaccess.thecvf.com/content_ECCV_2018/html/Rahaf_Aljundi_Memory_Aware_Synapses_ECCV_2018_paper.html},
    urldate = {2025-10-21},
    author = {Aljundi, Rahaf and Babiloni, Francesca and Elhoseiny, Mohamed and Rohrbach, Marcus and Tuytelaars, Tinne},
    year = {2018},
    doi = {https://doi.org/10.48550/arXiv.1711.09601},
    pages = {139--154},
}

@inproceedings{chaudhry_riemannian_2018,
    title = {Riemannian {Walk} for {Incremental} {Learning}: {Understanding} {Forgetting} and {Intransigence}},
    shorttitle = {Riemannian {Walk} for {Incremental} {Learning}},
    url = {https://openaccess.thecvf.com/content_ECCV_2018/html/Arslan_Chaudhry__Riemannian_Walk_ECCV_2018_paper.html},
    urldate = {2025-10-21},
    author = {Chaudhry, Arslan and Dokania, Puneet K. and Ajanthan, Thalaiyasingam and Torr, Philip H. S.},
    year = {2018},
    doi = {https://doi.org/10.48550/arXiv.1801.10112},
    pages = {532--547},
}

@inproceedings{benzing_unifying_2022,
    title = {Unifying {Importance} {Based} {Regularisation} {Methods} for {Continual} {Learning}},
    url = {https://proceedings.mlr.press/v151/benzing22a.html},
    language = {en},
    urldate = {2025-10-21},
    booktitle = {Proceedings of {The} 25th {International} {Conference} on {Artificial} {Intelligence} and {Statistics}},
    publisher = {PMLR},
    author = {Benzing, Frederik},
    month = may,
    year = {2022},
    note = {ISSN: 2640-3498},
    pages = {2372--2396},
}

@inproceedings{liu_rotate_2018,
    title = {Rotate your {Networks}: {Better} {Weight} {Consolidation} and {Less} {Catastrophic} {Forgetting}},
    shorttitle = {Rotate your {Networks}},
    url = {https://ieeexplore.ieee.org/abstract/document/8545895},
    doi = {10.1109/ICPR.2018.8545895},
    urldate = {2025-10-21},
    booktitle = {2018 24th {International} {Conference} on {Pattern} {Recognition} ({ICPR})},
    author = {Liu, Xialei and Masana, Marc and Herranz, Luis and van de Weijer, Joost and López, Antonio M. and Bagdanov, Andrew D.},
    month = aug,
    year = {2018},
    note = {ISSN: 1051-4651},
    pages = {2262--2268},
}

@inproceedings{lee_continual_2020,
    title = {Continual {Learning} {With} {Extended} {Kronecker}-{Factored} {Approximate} {Curvature}},
    url = {https://openaccess.thecvf.com/content_CVPR_2020/html/Lee_Continual_Learning_With_Extended_Kronecker-Factored_Approximate_Curvature_CVPR_2020_paper.html},
    urldate = {2025-10-21},
    author = {Lee, Janghyeon and Hong, Hyeong Gwon and Joo, Donggyu and Kim, Junmo},
    year = {2020},
    doi = {https://doi.org/10.48550/arXiv.2004.07507},
    pages = {9001--9010},
}

@inproceedings{park_continual_2019,
    title = {Continual {Learning} by {Asymmetric} {Loss} {Approximation} {With} {Single}-{Side} {Overestimation}},
    url = {https://openaccess.thecvf.com/content_ICCV_2019/html/Park_Continual_Learning_by_Asymmetric_Loss_Approximation_With_Single-Side_Overestimation_ICCV_2019_paper.html},
    urldate = {2025-10-21},
    author = {Park, Dongmin and Hong, Seokil and Han, Bohyung and Lee, Kyoung Mu},
    year = {2019},
    doi = {https://doi.org/10.48550/arXiv.1908.02984},
    pages = {3335--3344},
}

@article{hadsell_embracing_2020,
    title = {Embracing {Change}: {Continual} {Learning} in {Deep} {Neural} {Networks}},
    volume = {24},
    issn = {1364-6613, 1879-307X},
    shorttitle = {Embracing {Change}},
    url = {https://www.cell.com/trends/cognitive-sciences/abstract/S1364-6613(20)30219-9},
    doi = {10.1016/j.tics.2020.09.004},
    language = {English},
    number = {12},
    urldate = {2025-10-30},
    journal = {Trends in Cognitive Sciences},
    author = {Hadsell, Raia and Rao, Dushyant and Rusu, Andrei A. and Pascanu, Razvan},
    month = dec,
    year = {2020},
    pmid = {33158755},
    note = {Publisher: Elsevier},
    pages = {1028--1040},
}

@article{van_de_ven_three_2022,
    title = {Three types of incremental learning},
    volume = {4},
    copyright = {2022 The Author(s)},
    issn = {2522-5839},
    url = {https://www.nature.com/articles/s42256-022-00568-3},
    doi = {10.1038/s42256-022-00568-3},
    language = {en},
    number = {12},
    urldate = {2025-08-11},
    journal = {Nature Machine Intelligence},
    author = {van de Ven, Gido M. and Tuytelaars, Tinne and Tolias, Andreas S.},
    month = dec,
    year = {2022},
    note = {Publisher: Nature Publishing Group},
    pages = {1185--1197},
}

@article{scoville_loss_1957,
    title = {Loss of recent memory after bilateral hippocampal lesions},
    volume = {20},
    number = {11},
    journal = {J Neurol Neurosurg Psychiat},
    author = {Scoville, William Beecher and Milner, Brenda},
    year = {1957},
    doi = {https://doi.org/10.1136/jnnp.20.1.11},
    pages = {11--21},
}

\end{multicols}
\end{document}